\documentclass[12pt]{article}
\pdfoutput=1

\usepackage{amsmath}
\usepackage{amssymb}
\usepackage{epsfig}
\usepackage{epstopdf}
\usepackage{latexsym}
\usepackage{color}
\numberwithin{equation}{section}
\usepackage[
      colorlinks=false,
      linkcolor=darkblue,  
      urlcolor=blue,    
      filecolor=blue,     
      citecolor=red,
linktocpage=true,
      pdfstartview=FitV,
      bookmarksopen=true    
      ]{hyperref}

\usepackage{amssymb}
\begin{document}
%%%%%%%%%%%%%%%%%%%%%%%%%%%%%%%%%%%%%

%%%%%%%%%%%%%%%%%%%%%%%%%%%%%%%%%%%%%
\begin{titlepage}
%%%%%%%%%%%%%%%%%%%%%%%%%%%%%%%%%%%%%

\centerline
\centerline
\centerline
\bigskip
\centerline{\Huge \rm D4-branes wrapped on a topological disk} 
\bigskip
\bigskip
\bigskip
\bigskip
\bigskip
\bigskip
\bigskip
\bigskip
\centerline{\rm Minwoo Suh}
\bigskip
\centerline{\it Department of Physics, Kyungpook National University, Daegu 41566, Korea}
\bigskip
\centerline{\tt minwoosuh1@gmail.com} 
\bigskip
\bigskip
\bigskip
\bigskip
\bigskip
\bigskip
\bigskip

\begin{abstract}
\noindent Employing the method applied to M5-branes recently by Bah, Bonetti, Minasian and Nardoni, we study D4-branes wrapped on a disk with a non-trivial holonomy at the boundary. In $F(4)$ gauged supergravity in six dimensions, we find supersymmetric $AdS_4$ solutions and uplift the solutions to massive type IIA supergravity. We calculate the holographic free energy of dual three-dimensional superconformal field theories.
\end{abstract}

\vskip 7cm

\flushleft {August, 2021}

%%%%%%%%%%%%%%%%%%%%%%%%%%%%%%%%%%%%%
\end{titlepage}
%%%%%%%%%%%%%%%%%%%%%%%%%%%%%%%%%%%%%

\tableofcontents

%%%%%%%%%%%%%%%%%%%%%%%%%%%%%%%%%%%%%
\section{Introduction}
%%%%%%%%%%%%%%%%%%%%%%%%%%%%%%%%%%%%%

To our understanding of the AdS/CFT correspondence, \cite{Maldacena:1997re}, topological twisting has been essential in field theory, \cite{Witten:1988ze, Bershadsky:1995vm, Bershadsky:1995qy}, and also in supergravity, \cite{Maldacena:2000mw}. Recently, new examples of the AdS/CFT correspondence beyond the topological twisting have been proposed. The first kind of such $AdS$ solutions are from branes wrapped on a spindle which is topologically a two-sphere with orbifold singularities at the poles. There are $AdS$ solutions from D3-branes, \cite{Ferrero:2020laf, Hosseini:2021fge, Boido:2021szx}, M2-branes, \cite{Ferrero:2020twa, Cassani:2021dwa}, and M5-branes, \cite{Ferrero:2021wvk}, wrapped on a spindle. The solutions from D3-branes were previously found in \cite{Cvetic:1999xp, Gauntlett:2006af, Kunduri:2006uh, Gauntlett:2006ns, Kunduri:2007qy} in various contexts, but the interpretation as a spindle solution and the AdS/CFT correspondence are newly proposed. 

The second kind of such $AdS$ solutions are from branes wrapped on a topological disk which is a disk with non-trivial $U(1)$ holonomies at the boundary. The $AdS_5$ solutions from M5-branes wrapped on a topological disk was found in \cite{Bah:2021mzw, Bah:2021hei}. The dual field theory was proposed to be the Argyres-Douglas theory, \cite{Argyres:1995jj}, from 6d $\mathcal{N}=(2,0)$ theories on a sphere with irregular punctures. The construction was soon applied to $AdS_3$ solutions from D3-branes and M5-branes in \cite{Suh:2021ifj} and $AdS_2$ solutions from M2-branes in \cite{Suh:2021hef}. See also \cite{Couzens:2021tnv} and \cite{Couzens:2021rlk} for spindle and disk solutions from D3- and M2-branes. To recapitulate, topological dics and spindle are not manifolds with constant curvature and the supersymmetry is not realized by topological twist. 

In this paper, we construct the D4-D8 brane system wrapped on a topological disk. Five-dimensional superconformal field theories were first discovered in \cite{Seiberg:1996bd, Intriligator:1997pq} and their gravity dual was proposed to be the D4-D8 brane system, \cite{Ferrara:1998gv}, and found to be supersymmetric $AdS_6\,\times_w\,S^4$ solution, \cite{Brandhuber:1999np}, of massive type IIA supergravity, \cite{Romans:1985tz}. The solution is also realized as the supersymmetric fixed point, \cite{Cvetic:1999un}, of $F(4)$ gauged supergravity in six dimensions, \cite{Romans:1985tw}.

From topological twisting in $F(4)$ gauged supergravity, D4-D8-brane system wrapped on supersymmetric two- and three-cycles was studied in \cite{Nunez:2001pt, Naka:2002jz} and also from massive type IIA supergravity in \cite{Bah:2018lyv}. The D4-D8-brane system wrapped on a supersymmetric four-cycle provides the horizon geometry of supersymmetric $AdS_6$ black holes, \cite{Suh:2018tul}, and the Bekenstein-Hawking entropy was shown to match the field theory calculation of topologically twisted index, \cite{Hosseini:2018uzp, Crichigno:2018adf}. See \cite{Kim:2019fsg} also for non-supersymmetric solutions. So far, solutions were obtained from D4-D8-branes wrapped on supersymmetric cycles with constant curvature.

In this paper, we study D4-D8-branes wrapped on a disk with non-trivial holonomy at the boundary. In particular, we construct supersymmetric $AdS_4$ solutions of $F(4)$ gauged supergravity and uplift the solutions to massive type IIA supergravity. The dual field theories are 3d $\mathcal{N}=1$ SCFTs that arise from the twist compactification of 5d $\mathcal{N}=1$ $USp(2N)$ superconformal gauge theories, \cite{Seiberg:1996bd, Intriligator:1997pq}, on a topological disk. We calculate the holographic free energy of dual 3d superconformal field theories.

In section 2, we review $F(4)$ gauged supergravity in six dimensions. In section 3, we construct supersymmetric $AdS_4$ solutions and uplift the solutions to massive type IIA supergravity. In section 4, we conclude and discuss some open questions. The equations of motion are relegated in appendix A. In appendix B we derive the metric of D4-D8-branes smeared over four directions.

\bigskip

\noindent {\bf Note added:} Some time after this work, the spindle solutions from D4-branes are obtained in \cite{Faedo:2021nub, Giri:2021xta}. In appendix C, we show that the topological disk solution we obtain, in fact, matches the local solution of spindle in \cite{Faedo:2021nub}.

%%%%%%%%%%%%%%%%%%%%%%%%%%%%%%%%%%%%%
\section{$F(4)$ gauged supergravity in six dimensions}
%%%%%%%%%%%%%%%%%%%%%%%%%%%%%%%%%%%%%

We review $SU(2)\times{U}(1)$-gauged $\mathcal{N}\,=\,4$ supergravity in six dimensions \cite{Romans:1985tw}. The bosonic field content consists of the metric, $g_{\mu\nu}$, a real scalar, $\phi$, an $SU(2)$ gauge field, $A^I_\mu$, $I\,=\,1,\,2,\,3$, a $U(1)$ gauge field, $\mathcal{A}_\mu$, and a two-form gauge potential, $B_{\mu\nu}$. The fermionic field content is gravitinos, $\psi_{\mu{i}}$, and dilatinos, $\chi_i$, $i\,=\,1,\,2$. The field strengths are defined by
\begin{align}
\mathcal{F}_{\mu\nu}\,=&\,\partial_\mu\mathcal{A}_\nu-\partial_\nu\mathcal{A}_\mu\,, \notag \\
F^I_{\mu\nu}\,=&\,\partial_\mu{A}^I_\nu-\partial_\nu{A}^I_\mu+g\epsilon^{IJK}A^J_\mu{A}^K_\nu\,, \notag \\
G_{\mu\nu\rho}\,=&\,3\partial_{[\mu}B_{\nu\rho]}\,, \notag \\
\mathcal{H}_{\mu\nu}\,=&\,\mathcal{F}_{\mu\nu}+mB_{\mu\nu}\,.
\end{align}
The bosonic Lagrangian is given by
\begin{align}
e^{-1}\mathcal{L}\,=\,&-\frac{1}{4}R+\frac{1}{2}\partial_\mu\phi\partial^\mu\phi+\frac{1}{8}\left(g^2e^{\sqrt{2}\phi}+4gme^{-\sqrt{2}\phi}-m^2e^{-3\sqrt{2}\phi}\right) \notag \\
&-\frac{1}{4}e^{-\sqrt{2}\phi}\left(\mathcal{H}_{\mu\nu}\mathcal{H}^{\mu\nu}+F^I_{\mu\nu}F^{I\mu\nu}\right)+\frac{1}{12}e^{2\sqrt{2}\phi}G_{\mu\nu\rho}G^{\mu\nu\rho} \notag \\
&-\frac{1}{8}\epsilon^{\mu\nu\rho\sigma\tau\kappa}B_{\mu\nu}\left(\mathcal{F}_{\rho\sigma}\mathcal{F}_{\tau\kappa}+mB_{\rho\sigma}\mathcal{F}_{\tau\kappa}+\frac{1}{3}m^2B_{\rho\sigma}B_{\tau\kappa}+F^I_{\rho\sigma}F^I_{\tau\kappa}\right)\,,
\end{align}
where $g$ is the $SU(2)$ gauge coupling constant and $m$ is the mass of the two-form gauge potential. The supersymmetry transformations of the fermionic fields are
\begin{align}
\delta\psi_{\mu{i}}\,=&\,\nabla_\mu\epsilon_i+gA^I_\mu(T^I)_i\,^j\epsilon_j-\frac{1}{8\sqrt{2}}\left(ge^{-\frac{\phi}{\sqrt{2}}}+me^{-\frac{3\phi}{\sqrt{2}}}\right)\gamma_\mu\gamma_7\epsilon_i \notag \\
&-\frac{1}{8\sqrt{2}}e^{-\frac{\phi}{\sqrt{2}}}\left(\mathcal{F}_{\nu\lambda}+mB_{\nu\lambda}\right)\left(\gamma_\mu\,^{\nu\lambda}-6\delta_\mu\,^\nu\gamma^\lambda\right)\epsilon_i \notag \\
&-\frac{1}{4\sqrt{2}}e^{-\frac{\phi}{\sqrt{2}}}F^I_{\nu\lambda}\left(\gamma_\mu\,^{\nu\lambda}-6\delta_\mu\,^\nu\gamma^\lambda\right)\gamma_7(T^I)_i\,^j\epsilon_j \notag \\
&-\frac{1}{24}e^{\sqrt{2}\phi}G_{\nu\lambda\rho}\gamma_7\gamma^{\nu\lambda\rho}\gamma_\mu\epsilon_i\,, \\
\delta\chi_i\,=&\,\frac{1}{\sqrt{2}}\gamma^\mu\partial_\mu\phi\epsilon_i+\frac{1}{4\sqrt{2}}\left(ge^{-\frac{\phi}{\sqrt{2}}}-3me^{-\frac{3\phi}{\sqrt{2}}}\right)\gamma_7\epsilon_i \notag \\
&+\frac{1}{4\sqrt{2}}e^{-\frac{\phi}{\sqrt{2}}}\left(\mathcal{F}_{\mu\nu}+mB_{\mu\nu}\right)\gamma^{\mu\nu}\epsilon_i \notag \\
&+\frac{1}{2\sqrt{2}}e^{-\frac{\phi}{\sqrt{2}}}F^I_{\mu\nu}\gamma^{\mu\nu}\gamma_7(T^I)_i\,^j\epsilon_j \notag \\
&-\frac{1}{12}e^{\sqrt{2}\phi}G_{\mu\nu\lambda}\gamma_7\gamma^{\mu\nu\lambda}\epsilon_i\,,
\end{align}
where $T^I$, $I$ = 1, 2, 3, are the $SU(2)$ left-invariant one-forms,
\begin{equation}
T^I\,=\,-\frac{i}{2}\sigma^I\,.
\end{equation}
Mostly minus signature is employed. For $g>0$ and $m>0$ the theory admits a supersymmetric $AdS_6$ fixed point when $g\,=\,3m$. At the supersymmetric $AdS_6$ fixed point, all the fields are vanishing except the $AdS_6$ metric.

%%%%%%%%%%%%%%%%%%%%%%%%%%%%%%%%%%%%%
\section{Supersymmetric $AdS_4$ solutions}
%%%%%%%%%%%%%%%%%%%%%%%%%%%%%%%%%%%%%

%%%%%%%%%%%%%%%%%%%%%%%%%%%%%%%%%%%%%
\subsection{Supersymmetry equations}
%%%%%%%%%%%%%%%%%%%%%%%%%%%%%%%%%%%%%

We consider the background,
\begin{equation} \label{ansatzz}
ds^2\,=\,f(r)ds_{AdS_4}^2-g_1(r)dr^2-g_2(r)d\theta^2\,,
\end{equation}
with the gauge fields,
\begin{equation}
A^1\,=\,A^2\,=\,0\,, \qquad A^3\,=\,A_\theta(r)d\theta\,,
\end{equation}
and the scalar field, $\phi\,=\,\phi(r)$. The gamma matrices are given by
\begin{equation}
\gamma^{\alpha}\,=\,\rho^{\alpha}\otimes\mathbb{I}_4\,, \qquad \gamma^{\hat{r}}\,=\,i\rho_*\otimes\sigma^1\,, \qquad \gamma^{\hat{\theta}}\,=\,i\rho_*\otimes\sigma^2\,,
\end{equation}
where $\alpha\,=\,0,\,1,\,2,\,3$ are four-dimensional flat indices and the hatted indices are flat indiced for the corresponding coordinates. $\rho^\alpha$ are four-dimensional gamma matrices with $\{\rho^\alpha,\rho^\beta\}\,=\,2\eta^{\alpha\beta}$ and $\sigma^{1,2,3}$ are the Pauli matrices. The four- and six-dimensional chirality matrices are defined to be, respectively,
\begin{equation}
\rho_*\,=\,i\rho^0\rho^1\rho^2\rho^3\,, \qquad \gamma_7\,=\,\pm\rho_*\otimes\sigma^3\,,
\end{equation}
The spinor is given by
\begin{equation}
\epsilon_i\,=\,n_i\vartheta\otimes\eta\,,
\end{equation}
where $\vartheta$ is a Killing spinor on $AdS_4$ and $\eta\,=\,\eta(r,\theta)$. The Killing spinors satisfy
\begin{equation}
\nabla_\alpha^{AdS_4}\vartheta\,=\,\frac{1}{2}s\rho_\alpha\rho_*\vartheta\,,
\end{equation}
where $s\,=\,\pm1$.

The supersymmetry equations are obtained by setting the supersymmetry variations of the
fermionic fields to zero. From the supersymmetry variations, we obtain
\begin{align}
0\,=\,&\pm{s}\frac{i}{2}\gamma^{\hat{r}\hat{\theta}}\gamma_7\epsilon_i+\frac{1}{4}\frac{f'}{f^{1/2}g_1^{1/2}}\gamma^{\hat{r}}\epsilon_i-\frac{1}{8\sqrt{2}}\left(ge^{\frac{\phi}{\sqrt{2}}}+me^{-\frac{3\phi}{\sqrt{2}}}\right)f^{1/2}\gamma_7\epsilon_i \notag \\
&-\frac{1}{4\sqrt{2}}e^{-\frac{\phi}{\sqrt{2}}}A'_\theta\frac{f^{1/2}}{g_1^{1/2}g_2^{1/2}}\gamma^{\hat{r}\hat{\theta}}\gamma_72\left(T^3\right)_i\,^j\epsilon_j\,, \notag \\
0\,=\,&\partial_r\epsilon_i+\frac{1}{8\sqrt{2}}\left(ge^{\frac{\phi}{\sqrt{2}}}+me^{-\frac{3\phi}{\sqrt{2}}}\right)g_1^{1/2}\gamma^{\hat{r}}\gamma_7\epsilon_i+\frac{3}{4\sqrt{2}}e^{-\frac{\phi}{\sqrt{2}}}A'_\theta\frac{1}{g_2^{1/2}}\gamma^{\hat{\theta}}\gamma_72\left(T^3\right)_i\,^j\epsilon_j\,, \notag \\
0\,=\,&\partial_\theta\epsilon_i+\frac{1}{2}gA_\theta2\left(T^3\right)_i\,^j\epsilon_j+\frac{1}{4}\frac{g_2'}{g_1^{1/2}g_2^{1/2}}\gamma^{\hat{r}\hat{\theta}}\epsilon_i+\frac{1}{8\sqrt{2}}\left(ge^{\frac{\phi}{\sqrt{2}}}+me^{-\frac{3\phi}{\sqrt{2}}}\right)g_2^{1/2}\gamma^{\hat{\theta}}\gamma_7\epsilon_i \notag \\
&-\frac{3}{4\sqrt{2}}e^{-\frac{\phi}{\sqrt{2}}}A'_\theta\frac{1}{g_1^{1/2}}\gamma^{\hat{r}}\gamma_72\left(T^3\right)_i\,^j\epsilon_j\,, \notag \\
0\,=\,&\frac{1}{\sqrt{2}}\frac{1}{g_1^{1/2}}\phi'\gamma^{\hat{r}}\epsilon_i+\frac{1}{4\sqrt{2}}\left(ge^{\frac{\phi}{\sqrt{2}}}-3me^{-\frac{3\phi}{\sqrt{2}}}\right)\gamma_7\epsilon_i+\frac{1}{2\sqrt{2}}e^{-\frac{\phi}{\sqrt{2}}}A'_\theta\frac{1}{g_1^{1/2}g_2^{1/2}}\gamma^{\hat{r}\hat{\theta}}\gamma_72\left(T^3\right)_i\,^j\epsilon_j\,,
\end{align}
where the first three and the last equations are from the spin-3/2 and spin-1/2 field variations, respectively. By multiplying suitable functions and gamma matrices and adding the last equation to the first three equations, we obtain
\begin{align} \label{presusy}
0\,=\,&\pm{s}\frac{i}{2}\gamma^{\hat{r}\hat{\theta}}\gamma_7\epsilon_i+\frac{1}{2}g_1^{-1/2}f^{1/2}\left[\frac{1}{2}\frac{f'}{f}+\frac{1}{\sqrt{2}}\phi'\right]\gamma^{\hat{r}}\epsilon_i-\frac{m}{2\sqrt{2}}e^{-\frac{3\phi}{\sqrt{2}}}f^{1/2}\gamma_7\epsilon_i\,, \notag \\
0\,=\,&\partial_r\epsilon_i+\frac{1}{2\sqrt{2}}\phi'\epsilon_i+\frac{m}{2\sqrt{2}}e^{-\frac{3\phi}{\sqrt{2}}}g_1^{1/2}\gamma^{\hat{r}}\gamma_7\epsilon_i+\frac{1}{\sqrt{2}}g_2^{-1/2}e^{-\frac{\phi}{\sqrt{2}}}A'_\theta\gamma^{\hat{\theta}}2\left(T^3\right)_i\,^j\epsilon_j\,, \notag \\
0\,=\,&\partial_\theta\epsilon_i+\frac{1}{2}gA_\theta2\left(T^3\right)_i\,^j\epsilon_j-\frac{1}{\sqrt{2}}g_1^{-1/2}e^{-\frac{\phi}{\sqrt{2}}}A'_\theta\gamma^{\hat{r}}2\left(T^3\right)_i\,^j\epsilon_j+\frac{1}{2}g_1^{-1/2}g_2^{1/2}\left[\frac{1}{2}\frac{g_2'}{g_2}+\frac{1}{\sqrt{2}}\phi'\right]\gamma^{\hat{r}\hat{\theta}}\epsilon_i \notag \\
&+\frac{m}{2\sqrt{2}}e^{-\frac{3\phi}{\sqrt{2}}}g_2^{1/2}\gamma^{\hat{\theta}}\gamma_7\epsilon_i\,, \notag \\
0\,=\,&\frac{1}{4\sqrt{2}}\left(ge^{\frac{\phi}{\sqrt{2}}}-3me^{-\frac{3\phi}{\sqrt{2}}}\right)\gamma_7\epsilon_i+\frac{1}{\sqrt{2}}\phi'g_1^{-1/2}\gamma^{\hat{r}}\epsilon_i+\frac{1}{2\sqrt{2}}e^{-\frac{\phi}{\sqrt{2}}}A'_\theta{g}_1^{-1/2}g_2^{-1/2}\gamma^{\hat{r}\hat{\theta}}\gamma_72\left(T^3\right)_i\,^j\epsilon_j\,,
\end{align}
The spinor is supposed to have a charge under the $U(1)_\theta$ isometry,
\begin{equation}
\eta(r,\theta)\,=\,e^{in\theta}\widehat{\eta}(r)\,,
\end{equation}
where $n$ is a constant. It shows up in the supersymmetry equations in the form of $\left(-i\partial_\theta+\frac{1}{2}A_\theta\right)\eta\,=\,\left(n+\frac{1}{2}A_\theta\right)\eta$ which is invariant under
\begin{equation}
A^3\,\mapsto\,A^3-2\alpha_0d\theta\,, \qquad \eta\,\mapsto\,e^{i\alpha_0\theta}\eta\,,
\end{equation}
where $\alpha_0$ is a constant. We also define
\begin{equation} \label{AAhat}
\frac{1}{2}\widehat{A}_\theta\,=\,n+\frac{1}{2}A_\theta\,.
\end{equation}
We solve the equation of motion for the gauge fields and obtain
\begin{equation} \label{solgaugep}
A_\theta'\,=\,b\,e^{\sqrt{2}\phi}g_1^{1/2}g_2^{1/2}f^{-2}\,,
\end{equation}
where $b$ is a constant. Employing the expressions we discussed in \eqref{presusy} beside the second equation, we finally obtain the supersymmetry equations,
\begin{align}
0\,=\,&-isf^{-1/2}\eta+g_1^{-1/2}\left[\frac{1}{2}\frac{f'}{f}+\frac{1}{\sqrt{2}}\phi'\right]\left(\sigma^1\eta\right)\pm{i}\frac{m}{\sqrt{2}}e^{-\frac{3\phi}{\sqrt{2}}}\left(\sigma^3\eta\right)\,, \notag \\
0\,=\,&gg_2^{-1/2}\widehat{A}_\theta\left(\sigma^1\eta\right)\mp{i}\sqrt{2}f^{-2}e^{\frac{\phi}{\sqrt{2}}}\,b\,\left(\sigma^3\eta\right)+g_1^{-1/2}\left[\frac{1}{2}\frac{g_2'}{g_2}+\frac{1}{\sqrt{2}}\phi'\right]\left(i\sigma^2\eta\right)\pm{i}\frac{m}{\sqrt{2}}e^{-\frac{3\phi}{\sqrt{2}}}\eta\,, \notag \\
0\,=\,&\mp{i}\frac{1}{2\sqrt{2}}\left(ge^{\frac{\phi}{\sqrt{2}}}-3me^{-\frac{3\phi}{\sqrt{2}}}\right)\eta+\sqrt{2}g_1^{-1/2}\phi'\left(i\sigma^2\eta\right)\mp{i}\frac{1}{\sqrt{2}}f^{-2}e^{\frac{\phi}{\sqrt{2}}}\,b\,\left(\sigma^3\eta\right)\,.
\end{align}

The supersymmetry equations are in the form of $M^{(i)}\eta\,=\,0$, $i\,=\,1,\,2,\,3$, where $M^{(i)}$ are three $2\times{2}$ matrices, as we follow \cite{Bah:2021mzw, Bah:2021hei},
\begin{equation}
M^{(i)}\,=\,X_0^{(i)}\mathbb{I}_2+X_1^{(i)}\sigma^1+X_2^{(i)}\left(i\sigma^2\right)+X_3^{(i)}\sigma^3\,.
\end{equation}
We rearrange the matrices to introduce $2\times{2}$ matrices,
\begin{equation}
\mathcal{A}^{ij}\,=\,\text{det}\left(v^{(i)}|w^{(j)}\right)\,, \qquad \mathcal{B}^{ij}\,=\,\text{det}\left(v^{(i)}|v^{(j)}\right)\,, \qquad \mathcal{C}^{ij}\,=\,\text{det}\left(w^{(i)}|w^{(j)}\right)\,,
\end{equation}
from the column vectors of
\begin{equation}
v^{(i)}\,=\,\left(
\begin{array}{l}
 X_1^{(i)}+X_2^{(i)} \\
 -X_0^{(i)}-X_3^{(i)}
\end{array}
\right)\,, \qquad
w^{(i)}\,=\,\left(
\begin{array}{l}
 X_0^{(i)}-X_3^{(i)} \\
 -X_1^{(i)}+X_2^{(i)}
\end{array}
\right)\,.
\end{equation}

From the vanishing of $\mathcal{A}^{ij}$, $\mathcal{B}^{ij}$ and $\mathcal{C}^{ij}$, necessary conditions for non-trivial solutions are obtained. From $\mathcal{A}^{ii}\,=\,0$, we find
\begin{align} \label{Adiag}
0\,=\,&-\frac{1}{f}-\frac{1}{4g_1}\left(\frac{f'}{f}+\sqrt{2}\phi'\right)^2+\frac{m^2e^{-3\sqrt{2}\phi}}{2}\,, \notag \\
0\,=\,&\frac{2\,b^2\,e^{\sqrt{2}\phi}}{f^4}+\frac{1}{4g_1}\left(\frac{g_2'}{g_2}+\sqrt{2}\phi'\right)^2-\frac{m^2e^{-3\sqrt{2}\phi}}{2}-\frac{g^2\widehat{A}_\theta^2}{g_2}\,, \notag \\
0\,=\,&\frac{2\left(\phi'\right)^2}{g_1}+\frac{b^2e^{\sqrt{2}\phi}}{2f^4}-\frac{1}{8}\left(g\,e^{\frac{\phi}{\sqrt{2}}}-3m\,e^{-\frac{3\phi}{\sqrt{2}}}\right)^2\,.
\end{align}
From $\mathcal{A}^{ij}+\mathcal{A}^{ji}\,=\,0$, we find
\begin{align} \label{Aminus}
0\,=\,&-\frac{2\,b\,m\,e^{-\frac{2\phi}{\sqrt{2}}}}{f^2}-\frac{\sqrt{2}s\,m\,e^{-\frac{3\phi}{\sqrt{2}}}}{\sqrt{f}}-\frac{g\widehat{A}_\theta}{\sqrt{g_1}\sqrt{g_2}}\left(\frac{f'}{f}+\sqrt{2}\phi'\right)\,, \notag \\
0\,=\,&-\frac{b\,m\,e^{-\frac{2\phi}{\sqrt{2}}}}{f^2}+\frac{s}{\sqrt{2}\sqrt{f}}\left(g\,e^{\frac{\phi}{\sqrt{2}}}-3m\,e^{-\frac{3\phi}{\sqrt{2}}}\right)\,, \notag \\
0\,=\,&\frac{2\,b^2\,e^{\frac{2\phi}{\sqrt{2}}}}{f^4}+\frac{\sqrt{2}\phi'}{g_1}\left(\frac{g_2'}{g_2}+\sqrt{2}\phi'\right)+\frac{m\,e^{-\frac{3\phi}{\sqrt{2}}}}{2}\left(g\,e^{\frac{\phi}{\sqrt{2}}}-3m\,e^{-\frac{3\phi}{\sqrt{2}}}\right)\,.
\end{align}
From $\mathcal{A}^{ij}-\mathcal{A}^{ji}\,=\,0$, we find
\begin{align}
0\,=&\,\frac{1}{2g_1}\left(\frac{f'}{f}+\sqrt{2}\phi'\right)\left(\frac{g_2'}{g_2}+\sqrt{2}\phi'\right)-\frac{2\sqrt{2}s\,b\,e^{\frac{\phi}{\sqrt{2}}}}{f^{5/2}}-m^2e^{-3\sqrt{2}\phi}\,, \notag \\
0\,=&\,\frac{\sqrt{2}s\,b\,e^{\frac{\phi}{\sqrt{2}}}}{f^{5/2}}-\frac{\sqrt{2}}{g_1}\phi'\left(\frac{f'}{f}+\sqrt{2}\phi'\right)-\frac{m}{2}e^{-\frac{3\phi}{\sqrt{2}}}\left(g\,e^{\frac{\phi}{\sqrt{2}}}-3m\,e^{-\frac{3\phi}{\sqrt{2}}}\right)\,, \notag \\
0\,=&\,-\frac{m\,b\,e^{-\sqrt{2}\phi}}{f^2}-\frac{b\,e^{\frac{\phi}{\sqrt{2}}}}{f^2}\left(g\,e^{\frac{\phi}{\sqrt{2}}}-3m\,e^{-\frac{3\phi}{\sqrt{2}}}\right)+\frac{2\sqrt{2}\phi'\,g\,\widehat{A}_\theta}{\sqrt{g_1}\sqrt{g_2}}\,.
\end{align}
From $\mathcal{B}^{ij}+\mathcal{C}^{ij}\,=\,0$, we find 
\begin{align}
0\,=&\,-\frac{\sqrt{2}b\,e^{\frac{\phi}{\sqrt{2}}}}{f^2\sqrt{g_1}}\left(\frac{f'}{f}+\sqrt{2}\phi'\right)-\frac{s}{\sqrt{f}\sqrt{g_1}}\left(\frac{g_2'}{g_2}+\sqrt{2}\phi'\right)-\frac{\sqrt{2}e^{-\frac{3\phi}{\sqrt{2}}}g\,m\,\widehat{A}_\theta}{\sqrt{g_2}}\,, \notag \\
0\,=&\,\frac{b\,e^{\frac{\phi}{\sqrt{2}}}}{\sqrt{2}f^2\sqrt{g_1}}\left(\frac{f'}{f}+\sqrt{2}\phi'\right)+\frac{4\,s\,\phi'}{\sqrt{2}\sqrt{f}\sqrt{g_1}}\,, \notag \\
0\,=&\,-\frac{1}{2\sqrt{2}\sqrt{g_1}}\left(\frac{g_2'}{g_2}+\sqrt{2}\phi'\right)\left(g\,e^{\frac{\phi}{\sqrt{2}}}-3m\,e^{-\frac{3\phi}{\sqrt{2}}}\right)-\frac{2m\,e^{-\frac{3\phi}{\sqrt{2}}}\phi'}{\sqrt{g_1}}-\frac{\sqrt{2}b\,g\,e^{\frac{\phi}{\sqrt{2}}}\widehat{A}_\theta}{f^2\sqrt{g_2}}\,.
\end{align}
From $\mathcal{B}^{ij}-\mathcal{C}^{ij}\,=\,0$, we find 
\begin{align} \label{Abc}
0\,=&\,\frac{m\,e^{-\frac{3\phi}{\sqrt{2}}}}{\sqrt{2}\sqrt{g_1}}\left(\frac{f'}{f}+\sqrt{2}\phi'\right)-\frac{m\,e^{-\frac{3\phi}{\sqrt{2}}}}{\sqrt{2}\sqrt{g_1}}\left(\frac{g_2'}{g_2}+\sqrt{2}\phi'\right)-\frac{2s\,g\,\widehat{A}_\theta}{\sqrt{f}\sqrt{g_2}}\,, \notag \\
0\,=&\,\frac{1}{2\sqrt{2}\sqrt{g_1}}\left(\frac{f'}{f}+\sqrt{2}\phi'\right)\left(g\,e^{\frac{\phi}{\sqrt{2}}}-3m\,e^{-\frac{3\phi}{\sqrt{2}}}\right)+\frac{2m\,e^{-\frac{3\phi}{\sqrt{2}}}\phi'}{\sqrt{g_1}}\,, \notag \\
0\,=&\,-\frac{b\,e^{\frac{\phi}{\sqrt{2}}}}{\sqrt{2}f^2\sqrt{g_1}}\left(\frac{g_2'}{g_2}+\sqrt{2}\phi'\right)+\frac{4b\,e^{\frac{\phi}{\sqrt{2}}}\phi'}{f^2\sqrt{g_1}}-\frac{g\widehat{A}_\theta}{\sqrt{2}\sqrt{g_1}}\left(g\,e^{\frac{\phi}{\sqrt{2}}}-3m\,e^{-\frac{3\phi}{\sqrt{2}}}\right)\,.
\end{align}

%%%%%%%%%%%%%%%%%%%%%%%%%%%%%%%%%%%%%
\subsection{Supersymmetric solutions}
%%%%%%%%%%%%%%%%%%%%%%%%%%%%%%%%%%%%%

From the second equation of \eqref{Aminus}, we obtain
\begin{equation} \label{fsol}
f\,=\,\frac{2^{1/3}b^{2/3}m^{2/3}}{e^{\frac{4\phi}{3\sqrt{2}}}\left(s\left(ge^{\frac{\phi}{\sqrt{2}}}-3me^{-\frac{3\phi}{\sqrt{2}}}\right)\right)^{2/3}}\,,
\end{equation}
Then, from the third equation of \eqref{Adiag} with \eqref{fsol}, we obtain
\begin{equation}
g_1\,=\,\frac{16\,2^{1/3}b^{2/3}m^{8/3}\left(\phi'\right)^2}{\left(s\left(ge^{\frac{\phi}{\sqrt{2}}}-3me^{-\frac{3\phi}{\sqrt{2}}}\right)\right)^{2/3}\left(2^{1/3}b^{2/3}m^{8/3}-2e^{\frac{22\phi}{3\sqrt{2}}}\left(s\left(ge^{\frac{\phi}{\sqrt{2}}}-3me^{-\frac{3\phi}{\sqrt{2}}}\right)\right)^{2/3}\right)}\,,
\end{equation}
From the third equation of \eqref{Aminus}, we find an expression for $\sqrt{g_1}\sqrt{g_2}$,
\begin{equation} \label{g1g21}
\sqrt{g_1}\sqrt{g_2}\,=\,\frac{2\sqrt{2}gf^2\widehat{A}_\theta\phi'}{b\,e^{\frac{\phi}{\sqrt{2}}}\left(g\,e^{\frac{\phi}{\sqrt{2}}}-2me^{-\frac{3\phi}{\sqrt{2}}}\right)}\,.
\end{equation}
Also from \eqref{solgaugep}, we find another expression for $\sqrt{g_1}\sqrt{g_2}$,
\begin{equation} \label{g1g22}
\sqrt{g_1}\sqrt{g_2}\,=\,\frac{e^{-\sqrt{2}\phi}f^2\widehat{A}'_\theta}{b}\,.
\end{equation}
Equating \eqref{g1g21} and \eqref{g1g22}, we find an ordinary differential equation for $\widehat{A}_\theta$ and it gives
\begin{equation}
\widehat{A}_\theta\,=\,\mathcal{C}e^{\frac{3\phi}{\sqrt{2}}}\left(g\,e^{\frac{\phi}{\sqrt{2}}}-2me^{-\frac{3\phi}{\sqrt{2}}}\right)\,,
\end{equation}
where $\mathcal{C}$ is a constant. From \eqref{AAhat}, we find
\begin{equation}
A_\theta\,=\,\mathcal{C}e^{\frac{3\phi}{\sqrt{2}}}\left(g\,e^{\frac{\phi}{\sqrt{2}}}-2me^{-\frac{3\phi}{\sqrt{2}}}\right)+n\,.
\end{equation}
Then, from \eqref{g1g21} or \eqref{g1g22}, we obtain 
\begin{equation}
g_2\,=\,\frac{\mathcal{C}^2g^2\left(2^{1/3}b^{2/3}m^{8/3}-2e^{\frac{22\phi}{3\sqrt{2}}}\left(s\left(ge^{\frac{\phi}{\sqrt{2}}}-3me^{-\frac{3\phi}{\sqrt{2}}}\right)\right)^{2/3}\right)}{e^{\frac{4\phi}{3\sqrt{2}}}\left(s\left(ge^{\frac{\phi}{\sqrt{2}}}-3me^{-\frac{3\phi}{\sqrt{2}}}\right)\right)^{2/3}}\,.
\end{equation}
Therefore, we have determined all functions in terms of the scalar field, $\phi(r)$, and its derivative. The solution satisfies all the supersymmetry equations in \eqref{Adiag} to \eqref{Abc} and the equations motion which we present in appendix A. We can determine the scalar field by fixing the ambiguity in reparametrization of $r$ due to the covariance of the supersymmetry equations,
\begin{equation}
\phi(r)\,=\,\sqrt{2}\log{r}\,,
\end{equation}
where $r\,>\,0$.

Finally, let us summarize the solution. The metric is given by
\begin{equation} \label{metmet}
ds^2\,=\,\frac{B\,r^{2/3}}{m^2\left(s\left(gr^4-3m\right)\right)^{2/3}}\left[ds_{AdS_4}^2-\frac{32m^2r^{10/3}}{h(r)\left(s\left(gr^4-3m\right)\right)^{4/3}}dr^2-\frac{\mathcal{C}^2g^2m^2h(r)}{B}d\theta^2\right]\,,
\end{equation}
where we define
\begin{equation}
h(r)\,=\,B-2r^{16/3}\left(s\left(gr^4-3m\right)\right)^{2/3}\,,
\end{equation}
with
\begin{equation}
B\,=\,2^{1/3}b^{2/3}m^{8/3}\,.
\end{equation}
The gauge field is given by
\begin{equation}
\widehat{A}_\theta\,=\,\mathcal{C}\left(gr^4-2m\right).
\end{equation}
The metric can also be written as
\begin{equation}
ds^2\,=\,\frac{B\,r^{2/3}}{m^2\left(s\left(gr^4-3m\right)\right)^{2/3}}ds^2_{AdS_4}-\frac{32B\,r^4}{h(r)\left(s\left(gr^4-3m\right)\right)^2}dr^2-\frac{\mathcal{C}^2g^2r^{2/3}h(r)}{\left(s\left(gr^4-3m\right)\right)^{2/3}}d\theta^2\,.
\end{equation}

Now we consider the range of $r$ for regular solutions, $i.e.$, the metric functions are positive definite and the scalar fields are real.  We find regular solutions when we have
\begin{equation} \label{regrange}
0<r<r_1\,,
\end{equation}
where
\begin{align} \label{rone}
r_1^4\,=&\,\frac{m}{g}\left(1+X+X^{-1}\right)\,, \notag \\
X\,\equiv&\,2^{-\frac{2}{3}}e^{-\frac{2\pi{i}}{3}}\left(4+x+\sqrt{x\left(8+x\right)}\right)^{\frac{1}{3}}\,, \notag \\
x\,\equiv&\,bg^2m\,,
\end{align}
and $r_1$ is determined from $h(r_1)\,=\,0$. We plot a representative solution with $s=-1$, $b\,=\,0.1$, $\mathcal{C}\,=\,1$ and $g\,=\,3m\,=3$ in Figure 1. The metric on the space spanned by $\Sigma(r,\theta)$ in \eqref{metmet} has a topology of disk with the origin at $r\,=\,r_1$ and the boundary at $r\,=\,0$.

\begin{figure}[t]
\begin{center}
\includegraphics[width=2.0in]{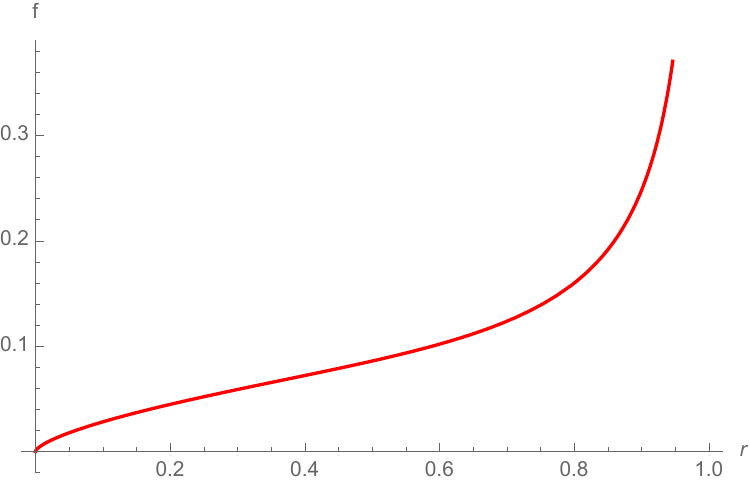} \qquad \includegraphics[width=2.0in]{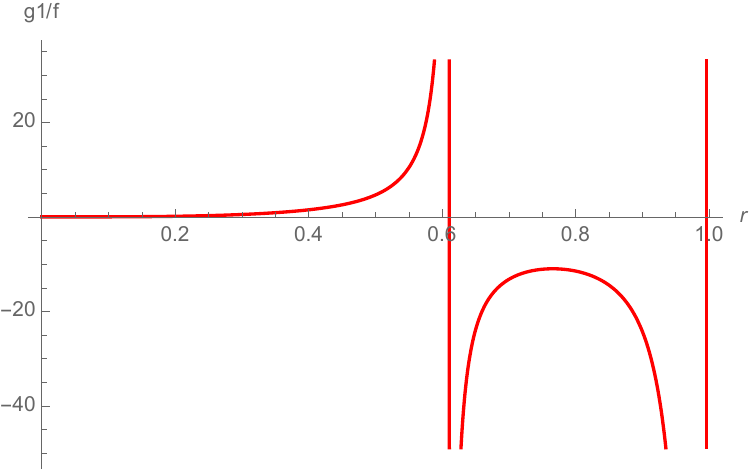} \qquad \includegraphics[width=2.0in]{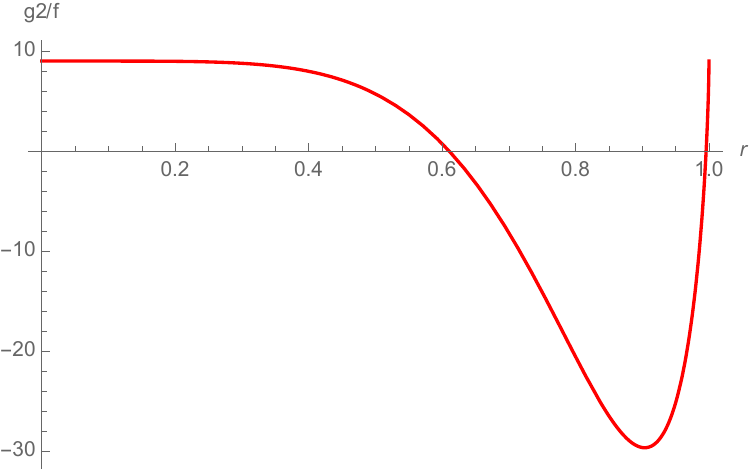}
\caption{{\it A representative solution with $s=-1$, $b=0.1$, $\mathcal{C}\,=\,1$ and $g\,=\,3m\,=3$. The solution is regular in the range of $0\,<\,r\,<\,r_1\,=\,0.610$.}}
\end{center}
\end{figure}

Near $r\,\rightarrow\,0$ the $AdS_4$ warp factor vanishes and it is a curvature singularity of the metric,
\begin{equation} \label{wsingu}
ds^2\,\approx\,\frac{Br^{2/3}}{3^{2/3}m^{8/3}}\left[ds_{AdS_4}^2-\frac{32m^{2/3}r^{10/3}}{3^{4/3}B}dr^2-\mathcal{C}^2g^2m^2d\theta^2\right]\,.
\end{equation}
This singularity is resolved when the solution is uplifted to massive type IIA supergravity.

Approaching $r\,=\,r_1$, the metric becomes to be 
\begin{equation}
ds^2\,=\,\frac{B\,r^{2/3}}{m^2\left(s\left(gr^4-3m\right)\right)^{2/3}}\left[ds_{AdS_4}^2-\frac{128m^2r^{10/3}\left[d\rho^2+\mathcal{C}^2\mathcal{E}^2(b;g,m)\rho^2d\theta^2\right]}{-h'(r_1)\left(s\left(gr^4-3m\right)\right)^{4/3}}\right]\,,
\end{equation}
where we introduced a new parametrization of coordinate, $\rho^2\,=\,r_1-r$. The function, $\mathcal{E}(b;g,m)$, is given by
\begin{equation} \label{functione}
\mathcal{E}(b;g,m)\,=\,\frac{2^{1/3}g}{b^{1/3}m^{4/3}}r_1^{8/3}\left(gr_1^4-2m\right)\left(gr_1^4-3m\right)^{1/3}\,.
\end{equation}
Then, the $\rho$-$\theta$ surface is locally an $\mathbb{R}^2/\mathbb{Z}_l$ orbifold if we set
\begin{equation} \label{toptwist}
\mathcal{C}\,=\,\frac{1}{l\mathcal{E}(b;g,m)}\,,
\end{equation}
where $l\,=\,1,\,2,\,3,\ldots\,\,$.

Employing the Gauss-Bonnet theorem, we calculate the Euler characteristic of $\Sigma$, the $r$-$\theta$ surface, from \eqref{metmet}. The boundary at $r\,=\,0$ is a geodesic and thus has vanishing geodesic curvature. The only contribution to the Euler characteristic is  
\begin{equation} \label{clb}
\chi\left(\Sigma\right)\,=\,\frac{1}{4\pi}\int_\Sigma{R}_\Sigma\text{vol}_\Sigma\,=\,\frac{2\pi}{4\pi}\frac{2^{4/3}\mathcal{C}g}{b^{1/3}m^{4/3}}r_1^{8/3}\left(gr_1^4-2m\right)\left(gr_1^4-3m\right)^{1/3}\,=\,\mathcal{C}\mathcal{E}(b;g,m)\,=\,\frac{1}{l}\,,
\end{equation}
where $0\,<\,\theta\,<2\pi$. This result is natural for a disk in an $\mathbb{R}^2/\mathbb{Z}_l$ orbifold centered at $r\,=\,r_1$ with $g\,=\,3m$.

%%%%%%%%%%%%%%%%%%%%%%%%%%%%%%%%%%%%%
\subsection{Uplift to massive type IIA supergravity}
%%%%%%%%%%%%%%%%%%%%%%%%%%%%%%%%%%%%%

We review the uplift formula of $F(4)$ gauged supergravity to massive type IIA supergravity, \cite{Cvetic:1999un}. We present the uplift formula in our conventions of \cite{Romans:1985tw}.{\footnote{In contrast to \cite{Romans:1985tw}, mostly plus signature is employed in \cite{Cvetic:1999un}. The couplings and fields in \cite{Cvetic:1999un} are related to the ones of \cite{Romans:1985tw} by
\begin{align}
g\,=&\,2\tilde{g}\,, \qquad X\,=\,e^{-\frac{\tilde{\phi}}{2\sqrt{2}}}\,=\,e^{\frac{\phi}{\sqrt{2}}}\,, \notag \\
A^I_\mu\,=&\,\frac{1}{2}\tilde{A}^I_\mu\,, \qquad A_\mu\,=\,\frac{1}{2}\tilde{A}_\mu\,, \qquad B_{\mu\nu}\,=\,\frac{1}{2}\tilde{B}_{\mu\nu}\,,
\end{align}
where the tilded ones are of \cite{Cvetic:1999un}. We will switch the solution to mostly plus signature, but keep following the normalization of \cite{Romans:1985tw}.}} The non-trivial fields are the metric, the dilaton, and the four-form flux, respectively,
\begin{align}
ds_{10}^2\,=&\,X^{1/8}\sin^{1/12}\xi\left(\Delta^{3/8}ds_6^2+\frac{8}{g^2}\Delta^{3/8}X^2d\xi^2+\frac{2}{g^2}\frac{\cos^2\xi}{\Delta^{5/8}X}ds_{\tilde{S}^3}^2\right)\,, \\
e^{\Phi}\,=&\,\frac{\Delta^{1/4}}{X^{5/4}\sin^{5/6}\xi}\,, \\ 
F_{(0)}\,=&\,m\,=\,\frac{g}{3}\,, \\
F_{(4)}\,=\,&-\frac{4\sqrt{2}}{3}\frac{U\sin^{1/3}\xi\cos^3\xi}{g^3\Delta^2}d\xi\wedge{vol}_{\tilde{S}^3}-8\sqrt{2}\frac{\sin^{4/3}\xi\cos^4\xi}{g^3\Delta^2X^3}dX\wedge{vol}_{\tilde{S}^3} \notag \\
&+\frac{8}{\sqrt{2}}\frac{\sin^{1/3}\xi\cos\xi}{g^2}F^I\wedge{h}^I\wedge{d}\xi-\frac{2}{\sqrt{2}}\frac{\sin^{4/3}\xi\cos^2\xi}{g^2\Delta{X}^3}F^I\wedge{h}^J\wedge{h}^K\epsilon_{IJK}\,.
\end{align}
We employ the metric and the volume form on the gauged three-sphere by
\begin{align}
ds_{\tilde{S}^3}^2\,=&\,\sum^3_{I=1}\left(\sigma^I-gA^I\right)^2\,, \notag \\
vol_{\tilde{S}^3}\,=&\,h_1\wedge{h_2}\wedge{h_3}\,,
\end{align}
where
\begin{align}
h^I\,=\,\sigma^I-gA^I\,,
\end{align}
and $\sigma^I$, $I\,=\,1,\,2,\,3$, are the $SU(2)$ left-invariant one-forms which satisfy
\begin{equation}
d\sigma^I\,=\,-\frac{1}{2}\epsilon_{IJK}\sigma^J\wedge\sigma^K\,.
\end{equation}
A choice of the left-invariant one-forms is
\begin{align}
\sigma^1\,=&\,-\sin\alpha_2\cos\alpha_3d\alpha_1+\sin\alpha_3d\alpha_2\,, \notag \\
 \sigma^2\,=&\,\sin\alpha_2\sin\alpha_3d\alpha_1+\cos\alpha_3d\alpha_2\,,  \notag \\
 \sigma^3\,=&\,\cos\alpha_2d\alpha_1+d\alpha_3\,.
\end{align}
We also defined quantities,
\begin{align}
X\,=&\,e^{\frac{\phi}{\sqrt{2}}}\,, \notag \\
\Delta\,=&\,X\cos^2\xi+X^{-3}\sin^2\xi\,, \notag \\
U\,=&\,X^{-6}\sin^2\xi-3X^2\cos^2\xi+4X^{-2}\cos^2\xi-6X^{-2}\,.
\end{align}

For our solutions, we have $X\,=\,r$. In particular, the metric can be written by
\begin{align} \label{upmet}
ds^2_{10}\,=&\,\frac{B\Delta^{3/8}r^{19/24}\sin^{1/12}\xi}{m^2\left(s\left(gr^4-3m\right)\right)^{2/3}}\left[ds_{AdS_4}^2+\frac{32m^2r^{10/3}}{h\left(s\left(gr^4-3m\right)\right)^{4/3}}dr^2+\frac{\mathcal{C}^2g^2m^2h}{B}d\theta^2\right. \notag \\
+&\left.\frac{8m^2r^2\left(s\left(gr^4-3m\right)\right)^{2/3}}{B\,g^2r^{2/3}}d\xi^2+\frac{2m^2\cos^2\xi\left(s\left(gr^4-3m\right)\right)^{2/3}}{B\,g^2r^{5/3}\Delta}ds^2_{\widetilde{S}^3}\right]\,,
\end{align}
where we have
\begin{equation}
\Delta\,=\,r\cos^2\xi+r^{-3}\sin^2\xi\,.
\end{equation}

%%%%%%%%%%%%%%%%%%%%%%%%%%%%%%%%%%%%%
\subsection{Uplifted metric}
%%%%%%%%%%%%%%%%%%%%%%%%%%%%%%%%%%%%%

The six-dimensional internal space of the uplifted metric is an $S_\theta^1\,\times\,S^3$ fibration over the 2d base space, $B_2$, of $(r,\xi)$. The 2d base space is a rectangle of $(r,\xi)$ over $[0,r_1)\,\times\left[0,\frac{\pi}{2}\right]$. See Figure 2. We explain the geometry of the internal space by three regions of the 2d base space, $B_2$.

\begin{itemize}
\item Region I: The side of $\mathsf{P}_1\mathsf{P}_2$.
\item Region II: The sides of $\mathsf{P}_2\mathsf{P}_3$ and $\mathsf{P}_3\mathsf{P}_4$.
\item Region III: The side of $\mathsf{P}_1\mathsf{P}_4$.
\end{itemize}

\begin{figure}[t]
\begin{center}
\includegraphics[width=4.5in]{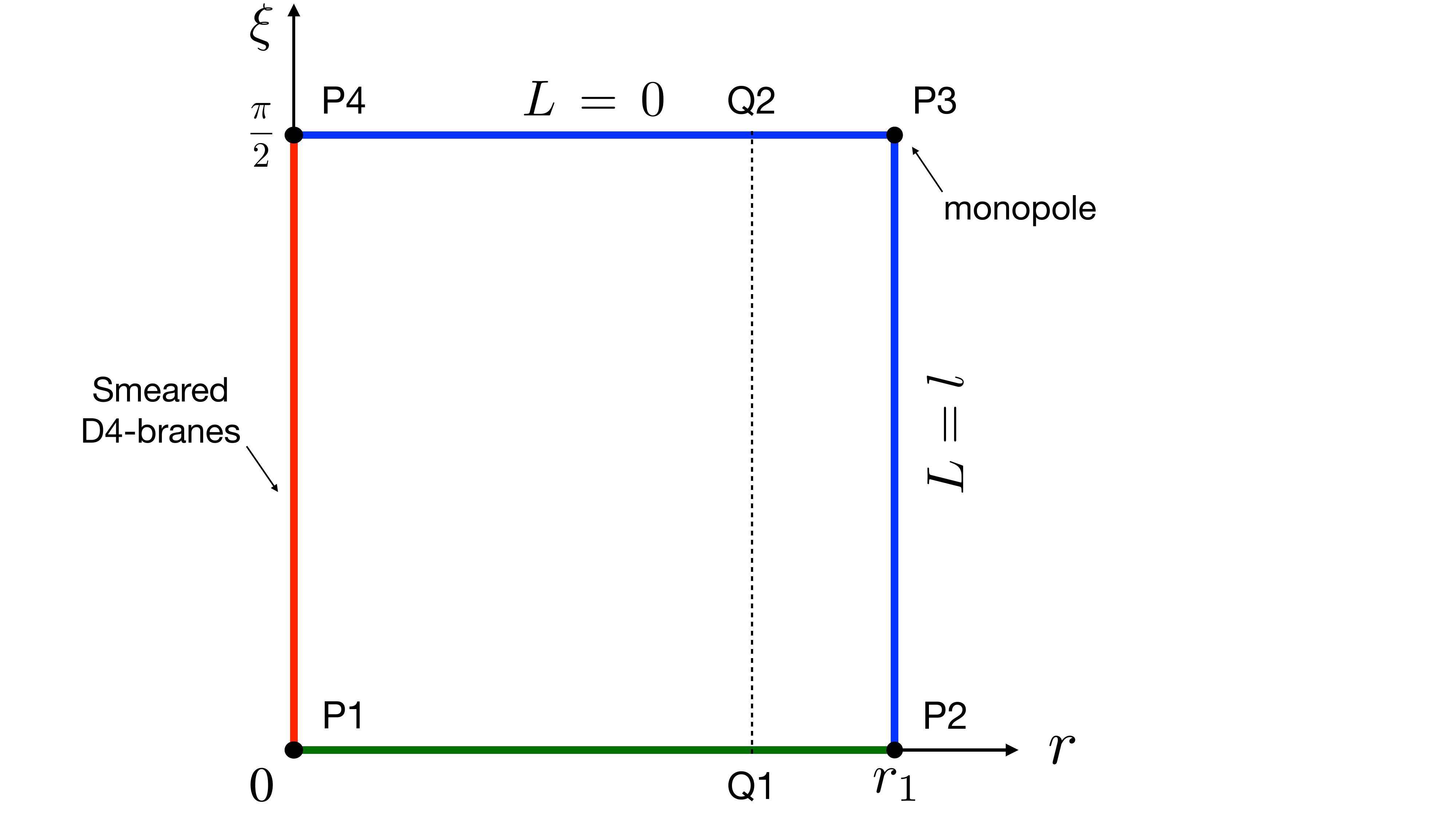}
\caption{{\it The two-dimensional base space, $B_2$, spanned by $r$ and $\xi$.}}
\end{center}
\end{figure}

\noindent {\bf Region I:} On the side of $\xi\,=\,0$, unlike the D3-, M2-, and M5-branes wrapped on a topological disk, there is no circle that shrinks.

\bigskip

\noindent {\bf Region II: Monopole} We break $\left(\sigma^3-gA^3\right)^2$ in $S^3$ and complete the square of $d\theta$, \cite{Bah:2021hei, Couzens:2021rlk}, to obtain the metric of
\begin{align}
ds^2_{10}\,=&\,\frac{B\Delta^{3/8}r^{19/24}\sin^{1/12}\xi}{m^2\left(s\left(gr^4-3m\right)\right)^{2/3}}\left[ds_{AdS_4}^2+\frac{32m^2r^{10/3}}{h\left(s\left(gr^4-3m\right)\right)^{4/3}}dr^2\right. \notag \\
+&\frac{8m^2r^2\left(s\left(gr^4-3m\right)\right)^{2/3}}{B\,g^2r^{2/3}}d\xi^2+\frac{2m^2\cos^2\xi\left(s\left(gr^4-3m\right)\right)^{2/3}}{B\,g^2r^{5/3}\Delta}\left((\sigma^1)^2+(\sigma^2)^2\right) \notag \\
+&R_\theta^2\left(d\theta-gL\sigma^3\right)^2+R_{\sigma^3}^2(\sigma^3)^2\Big]\,.
\end{align}
The metric functions are defined to be
\begin{align}
R_\theta^2\,=&\frac{\mathcal{C}^2m^2\left(g^4\Delta{h}r^{5/3}+2\left(gr^4-2m\right)^2\left(s\left(gr^4-3m\right)\right)^{2/3}\cos^2\xi\right)}{Bg^2\Delta{r}^{4/3}}\,, \notag \\
R_{\sigma^3}^2\,=&\,\frac{2g^2m^2h\left(s\left(gr^4-3m\right)\right)^{2/3}\cos^2\xi}{B\left(g^4\Delta{h}r^{5/3}+2\left(gr^4-2m\right)^2\left(s\left(gr^4-3m\right)\right)^{2/3}\cos^2\xi\right)}\,, \notag \\
L\,=&\,\frac{2\left(gr^4-2m\right)^2\left(s\left(gr^4-3m\right)\right)^{2/3}\cos^2\xi}{g\,\mathcal{C}\left(g^4\Delta{h}r^{5/3}+2\left(gr^4-2m\right)^2\left(s\left(gr^4-3m\right)\right)^{2/3}\cos^2\xi\right)}\,.
\end{align}

The function, $L(r,\xi)$, is piecewise constant along the sides of $r\,=\,r_1$ and $\xi\,=\,\frac{\pi}{2}$ of the 2d base, $B_2$,
\begin{equation}
L\left(r,\frac{\pi}{2}\right)\,=\,0\,, \qquad L\left(r_1,\xi\right)\,=\,\frac{1}{\mathcal{C}\mathcal{E}(b;g,m)}\,=\,l\,.
\end{equation}
The jump in $L$ at the corner, $(r,\xi)\,=\,\left(r_1,\frac{\pi}{2}\right)$, indicates the existence of a monopole source for the $D\theta$ fibration. Due to the complexity of the expressions, we were not able to take the $r\,\rightarrow\,r_1$ and $\xi\,\rightarrow\,\frac{\pi}{2}$ limit to obtain the metric of the monopole.

\bigskip

\noindent {\bf Region III: Smeared D4-D8-branes} The singularity at $r\,\rightarrow\,0$ in the warp factor of five-dimensional metric, \eqref{wsingu}, has been resolved in the uplifted metric, \eqref{upmet}. On the other hand, there is a singularity at $\left(r\,\rightarrow\,0,\,\sin\xi\rightarrow{0}\right)$ and we consider this singularity. We first transform the metric from Einstein frame to string frame, $ds_{\text{string}}^2=e^{\Phi/2}ds_{\text{Einstein}}^2$. Then, in the limit, the metric asymptotes to
\begin{align} \label{smeared48}
ds_{\text{string}}^2\,\approx&\,\frac{B}{3^{2/3}m^{8/13}}r^{-4/3}\sin^{2/3}\xi\left[ds_{AdS_4}^2+g^2m^2\mathcal{C}^2d\theta^2\right] \notag \\
&+\frac{8}{g^2}\sin^{2/3}\xi\left[d\xi^2+\frac{\cot^2\xi}{4}ds_{\tilde{S}^3}^2\right]+\frac{32}{9m^2}r^2\sin^{2/3}{\xi}dr^2\,.
\end{align}
The metric implies the smeared D4-D8-brane sources. The D4-D8-branes are 
\begin{itemize}
\item extended along the $AdS_4$ and $\theta$ directions;
\item localized at the center of $r$ direction;
\item smeared along the $\xi$ and $\tilde{S}^3$ directions. 
\end{itemize}
This matches the metric of D4-D8-branes smeared over four directions which can be obtained by following \cite{Bah:2017wxp}. We present the derivation of metric for smeared D4-D8-branes in appendix B.{\footnote{We would like to thank Hyojoong Kim for very helpful comments on this limit.}}

Lastly, we briefly present the comparison of our geometry with the geometry of wrapped M5-branes in \cite{Bah:2021mzw, Bah:2021hei}. The overall geometries are given by 
\begin{align} \label{compar}
\text{Wrapped D4-branes}: \qquad &AdS_4\,\times\,\emptyset\,\times\,S_\theta^1\,\times\,S^3\,\times\,[r,\xi]\,, \notag \\
\text{Wrapped M5-branes}: \qquad &AdS_5\,\times\,\,S^2\,\,\times\,\,S_z^1\,\times\,S_\phi^1(D\phi)\,\times\,[w,\mu]\,,
\end{align}
where $\emptyset$ is empty and we denote the gauged coordinates with $D$, $e.g.$, $D\phi$. For each metric, we presented the factors in the same order so that the corresponding factors are easily found.

%%%%%%%%%%%%%%%%%%%%%%%%%%%%%%%%%%%%%
\subsection{Flux quantization}
%%%%%%%%%%%%%%%%%%%%%%%%%%%%%%%%%%%%%

In order to properly quantize the flux fields in massive type IIA supergravity, we rescale the fields with a positive parameter, $\lambda$, \cite{Faedo:2021nub},
\begin{align}
d\hat{s}_{\text{string}}^2\,&=\,\lambda^2ds_{\text{string}}^2\,, \qquad e^{\hat{\Phi}}\,=\,\lambda^2e^\Phi\,, \qquad \hat{B}_{(2)}\,=\,\lambda^2B_{(2)}\,, \notag \\
\hat{F}_{(0)}\,&=\,\lambda^{-3}F_{(0)}\,, \qquad \hat{C}_{(n-1)}\,=\,\lambda^{n-3}C_{(n-1)}\,,
\end{align}
where the metric is in the string frame and $n=2,4$. Under the transformation the equations of motion are invariant. As there is only one free parameter, $g=m/3$, and two constraints from the quantizations of $F_{(0)}$ and $F_{(4)}$, the additional parameter, $\lambda$, is required. For the convenience of notation, we will remove the hat on the fields from now on.

The quantization condition on the Romans mass, $F_{(0)}\,=\,\frac{g}{3\lambda^3}$, is given by
\begin{equation} \label{f0n0}
\left(2\pi{l}_s\right)F_{(0)}\,\equiv\,n_0\,\in\,\mathbb{Z}\,,
\end{equation}
where $n_0\,=8-N_f$ and $N_f$ is the number of D8-branes. 

We consider the flux quantization condition for the four-form flux. The integral of the four-form flux over any four-cycle in the internal space is an integer, see, $e.g.$, \cite{Bah:2018lyv},
\begin{equation}
\frac{1}{\left(2\pi{l}_s\right)^3}\int_{M_4}F_{(4)}\,\in\,\mathbb{Z}\,,
\end{equation}
where $l_s$ is the string length.

First, we consider the $F_{(4)\,\xi\alpha_1\alpha_2\alpha_3}$ component of the four-form flux and we obtain
\begin{equation} \label{fluxq}
\frac{1}{\left(2\pi{l}_s\right)^3}\int{F}_{(4)\,\xi\alpha_1\alpha_2\alpha_3}\,=\,\frac{1}{\left(2\pi{l}_s\right)^3}\int\left(-\lambda\frac{4\sqrt{2}}{3}\frac{U\sin^{1/3}\xi\cos^3\xi}{g^2\Delta^2}\right)d\xi\wedge\text{vol}_{\widetilde{S}^3}\,=\,\frac{3\lambda}{2\sqrt{2}\pi{l}_s^3g^3}\,\equiv\,N\,,
\end{equation}
where $\text{vol}_{\tilde{S}^3}=2\pi^2$ and $N\,\in\,\mathbb{N}$ is the number of D4-D8-branes wrapping the two-dimensional manifold, $\Sigma$. This integration contour corresponds to the interval, $\mathsf{Q}_1\mathsf{Q}_2$ in Figure 2.

Second, we consider the $F_{(4)\,r\theta\alpha_3\xi}$ component of the four-form flux and we obtain
\begin{equation}
\frac{1}{\left(2\pi{l}_s\right)^3}\int{F}_{(4)\,r\theta\alpha_3\xi}\,=\,\frac{1}{\left(2\pi{l}_s\right)^3}\int\lambda\frac{8}{\sqrt{2}}\frac{\sin^{1/3}\xi\cos\xi}{g^2}F^I\wedge\,d\alpha_3\,\wedge\,d\xi\,=\,\frac{3\lambda\mathcal{C}r_1^4}{\sqrt{2}\pi{l}_s^3g}\,.
\end{equation}
Plugging $\mathcal{C}$ from \eqref{toptwist} and $l_s$ from \eqref{fluxq}, we obtain
\begin{align} \label{fluxk}
\frac{1}{\left(2\pi{l}_s\right)^3}\int{F}_{(4)\,r\theta\alpha_3\xi}\,=\,\frac{N}{l}\frac{2g^2r_1^4}{\mathcal{E}(b;g,m)}\,\equiv\,K\,
\end{align}
where $K\,\in\,\mathbb{Z}$ is another integer.

From \eqref{f0n0} and \eqref{fluxq}, we determine
\begin{equation}
g^8\,=\,\frac{1}{\left(2\pi{l}_s\right)^8}\frac{144\sqrt{2}\pi^6}{N^3n_0}\,, \qquad \lambda^8\,=\,\frac{2\sqrt{2}\pi^2}{9Nn_0^3}\,.
\end{equation}

By eliminating $\mathcal{E}(b)$ from the constraints, \eqref{toptwist} and \eqref{fluxk}, and with the expression of $r_1(b)$ in \eqref{rone}, we obtain{\footnote{We are very happy to acknowledge an anonymous referee who first derived \eqref{bone} and encouraged us to find explicit expressions of $\mathcal{C}$ and $b$ in terms of the quantum numbers, $N$ and $K$.}}
\begin{equation} \label{bone}
b\,=\,\frac{K^2\left(K-6gm\mathcal{C}N\right)}{4g^2m\left(gm\mathcal{C}N\right)^3}\,,
\end{equation}
and we also find
\begin{equation} \label{ronerone}
r_1(b)^4\,=\,\frac{K}{2g^2\mathcal{C}N}\,.
\end{equation}
Then, by plugging $r_1(b)$, \eqref{ronerone}, in \eqref{toptwist} with the expression of $\mathcal{E}(b)$ in \eqref{functione}, we also find another expression for $b$,
\begin{equation} \label{btwo}
b\,=\,\frac{K^2l^3\left(K-6gm\mathcal{C}N\right)\left(K-4gm\mathcal{C}N\right)^3}{32g^5m^4\mathcal{C}^3N^6}\,.
\end{equation}
Finally, identifying \eqref{bone} and \eqref{btwo}, we can solve for $\mathcal{C}$ and then for $b$ in terms of the quantum numbers, $N$ and $K$,
\begin{equation} \label{candb}
\mathcal{C}\,=\,\frac{Kl-2N}{4gmNl}\,, \qquad b\,=\,-\frac{8K^2l^2\left(Kl-6N\right)}{g^2m\left(Kl-2N\right)^3}\,.
\end{equation}

%%%%%%%%%%%%%%%%%%%%%%%%%%%%%%%%%%%%%
\subsection{Holographic free energy}
%%%%%%%%%%%%%%%%%%%%%%%%%%%%%%%%%%%%%

Now we calculate the holographic free energy of dual 3d SCFTs. Consider the metric of the form,
\begin{equation}
ds_{10}^2\,=\,e^{2\mathcal{A}}\left(ds_{AdS_4}^2+ds^2_{M_6}\right)\,,
\end{equation}
in the Einstein frame. The formula for holographic free energy is given in $e.g.$, \cite{Bah:2018lyv},
\begin{equation}
\mathcal{F}\,=\,\frac{16\pi^3\lambda^4}{\left(2\pi{l}_s\right)^8}\int_{M_6}e^{8\mathcal{A}-2\Phi}\text{vol}_{M_6}\,,
\end{equation}
where the ten-dimensional metric is in the string frame, $ds_{\text{string}}^2=e^{\Phi/2}ds_{\text{Einstein}}^2$. Employing the formula, we find
\begin{align} \label{cformula}
\mathcal{F}\,=&\,\frac{16\pi^3\lambda^4}{\left(2\pi{l}_s\right)^8}\int\frac{32\sqrt{2}B^{3/2}\mathcal{C}r^3\cos^3\xi\sin^{1/3}\xi}{g^3m^2\left(s\left(gr^4-3m\right)\right)^2}\,\text{vol}_{\tilde{S}^3}dr\,d\theta\,d\xi \notag \\
=&\,\frac{16\pi^3\lambda^4}{\left(2\pi{l}_s\right)^8}\left[\frac{36\,b\,\mathcal{C}\,m^2}{5g^4\left(s\left(gr^4-3m\right)\right)}\right]_{r_\text{min}}^{r_\text{max}}2\pi\,2\pi^2\,,
\end{align}
where we have $0<\theta<2\pi$, $0<\xi<\frac{\pi}{2}$, and $\text{vol}_{\tilde{S}^3}=2\pi^2$. For the solutions in \eqref{regrange} with $r_{\text{min}}\,=\,0$ and $r_{\text{max}}\,=\,r_1$, we obtain the holographic free energy,
\begin{equation}
\mathcal{F}\,=\,\frac{3\lambda^4}{5\pi^2l_s^8}\frac{b\,\mathcal{C}\,m\,r_1^4}{g^3\left(gr_1^4-3m\right)}\,=\,\frac{64\,\,2^{1/4}\pi{K}^3N^{3/2}l^2}{15\sqrt{8-N_f}\left(Kl-2N\right)^2}\,,
\end{equation}
where we used \eqref{candb}. If we set $K\sim{N}$, the free energy scales as $\mathcal{F}\sim{N}^{5/2}$ as the free energy of 5d SCFTs in \cite{Seiberg:1996bd, Intriligator:1997pq}. Even though the uplifted solutions have singularities, we obtain a well-defined finite result for free energy.

%%%%%%%%%%%%%%%%%%%%%%%%%%%%%%%%%%%%%
\section{Conclusions}
%%%%%%%%%%%%%%%%%%%%%%%%%%%%%%%%%%%%%

Employing the method applied to M5-branes recently by \cite{Bah:2021mzw, Bah:2021hei}, we constructed supersymmetric $AdS_4$ solutions from D4-D8-branes wrapped on a two-dimensional manifold with non-constant curvature. We uplifted the solutions to massive type IIA supergravity and calculated the holographic free energy of dual three-dimensional superconformal field theories.

The first natural question would be to identify the three-dimensional superconformal field theory which is dual to the solution and match the free energy calculated from the field theory.

In this work, we only constructed a class of $AdS_4$ fixed points from D4-D8-branes on a non-constant curvature manifold. The holographic RG flow from the $AdS_6$ fixed point dual to 5d superconformal field theories would enable us to understand more details of the solution.

From matter coupled $F(4)$ gauged supergravity, \cite{Andrianopoli:2001rs}, wrapped D4-D8-brane solutions on constant-curvature manifolds were previously studied in \cite{Karndumri:2015eta, Hosseini:2018usu, Suh:2018szn}. We would like to generalize our solutions in matter coupled $F(4)$ gauged supergravity. See also \cite{Hosseini:2020wag}.

Among twist compactifications of branes and their dual field theories, D4-D8-brane system is lesser understood and we look forward to seeing development to come in the future.

%%%%%%%%%%%%%%%%%%%%%%%%%%%%%%%%%%%%%
\bigskip
\leftline{\bf Acknowledgements}
\noindent We thank Chris Couzens and Hyojoong Kim for very helpful discussions. We also thank an anonymous referee for very instructive suggestions and encouragement.  This research was supported by the National Research Foundation of Korea under the grant NRF-2019R1I1A1A01060811.
%%%%%%%%%%%%%%%%%%%%%%%%%%%%%%%%%%%%%

%%%%%%%%%%%%%%%%%%%%%%%%%%%%%%%%%%%%%
\appendix
\section{The equations of motion}
\renewcommand{\theequation}{A.\arabic{equation}}
\setcounter{equation}{0} 
%%%%%%%%%%%%%%%%%%%%%%%%%%%%%%%%%%%%%

In this appendix, we present the equations of motion of $F(4)$ gauged supergravity,
\begin{align}
&R_{\mu\nu}\,=\,2\partial_\mu\phi\partial_\nu\phi+\frac{1}{8}g_{\mu\nu}\left(g^2e^{\sqrt{2}\phi}+4gme^{-\sqrt{2}\phi}-m^2e^{-3\sqrt{2}\phi}\right)-2e^{-\sqrt{2}\phi}\left(\mathcal{H}_\mu\,^\rho\mathcal{H}_{\nu\rho}-\frac{1}{8}g_{\mu\nu}\mathcal{H}_{\rho\sigma}\mathcal{H}^{\rho\sigma}\right) \notag \\ 
& \,\,\,\,\,\,\,\,\,\,\,\,\,\,\,\,\,\,\,\,\,\,\,\,\,\, -2e^{-\sqrt{2}\phi}\left(F^I_\mu\,^\rho{F}^I_{\nu\rho}-\frac{1}{8}g_{\mu\nu}F^I_{\rho\sigma}F^{I\rho\sigma}\right)+e^{2\sqrt{2}\phi}\left(G_\mu\,^{\rho\sigma}G_{\nu\rho_\sigma}-\frac{1}{6}g_{\mu\nu}G_{\rho\sigma\tau}G^{\rho\sigma\tau}\right)\,, \\
&\frac{1}{\sqrt{-g}}\partial_\mu\left(\sqrt{-g}g^{\mu\nu}\partial_\nu\phi\right)\,=\,\frac{1}{4\sqrt{2}}\left(g^2e^{\sqrt{2}\phi}-4gme^{-\sqrt{2}\phi}+3m^2e^{-3\sqrt{2}\phi}\right) \notag \\ 
& \,\,\,\,\,\,\,\,\,\,\,\,\,\,\,\,\,\,\,\,\,\,\,\,\,\, +\frac{1}{2\sqrt{2}}e^{-\sqrt{2}\phi}\left(\mathcal{H}_{\mu\nu}\mathcal{H}^{\mu\nu}+F^I_{\mu\nu}F^{I\mu\nu}\right)+\frac{1}{3\sqrt{2}}e^{2\sqrt{2}\phi}G_{\mu\nu\rho}G^{\mu\nu\rho}\,, \\
&\mathcal{D}_\nu\left(e^{-\sqrt{2}\phi}\mathcal{H}^{\nu\mu}\right)\,=\,\frac{1}{6}e\epsilon^{\mu\nu\rho\sigma\tau\kappa}\mathcal{H}_{\nu\rho}G_{\sigma\tau\kappa}\,, \\
&\mathcal{D}_\nu\left(e^{-\sqrt{2}\phi}F^{I\nu\mu}\right)\,=\,\frac{1}{6}e\epsilon^{\mu\nu\rho\sigma\tau\kappa}F^I_{\nu\rho}G_{\sigma\tau\kappa}\,, \\
&\mathcal{D}_\rho\left(e^{2\sqrt{2}\phi}G^{\rho\mu\nu}\right)\,=\,-\frac{1}{4}e\epsilon^{\mu\nu\rho\sigma\tau\kappa}\left(\mathcal{H}_{\rho\sigma}\mathcal{H}_{\tau\kappa}+F^I_{\rho\sigma}F^I_{\tau\kappa}\right)-me^{-\sqrt{2}\phi}\mathcal{H}^{\mu\nu}\,.
\end{align}

%%%%%%%%%%%%%%%%%%%%%%%%%%%%%%%%%%%%%
\section{Smeared D4-D8-branes}
\renewcommand{\theequation}{B.\arabic{equation}}
\setcounter{equation}{0} 
%%%%%%%%%%%%%%%%%%%%%%%%%%%%%%%%%%%%%

In this appendix, we derive the metric of D4-D8-brane system smeared over four directions by following appendix B of \cite{Bah:2017wxp}.

The metric describing D4-branes in the worldvolume of D8-branes were constructed in \cite{Youm:1999ti}. In particular, for the D4-branes extending from $x_0$ to $x_4$ and the D8-branes along all directions beside $x_9$, the metric is given by
\begin{equation}
ds^2\,=\,\left(H_8H_4\right)^{-1/2}\left(-dx_0^2+\cdots+dx_4^2\right)+H_4^{1/2}H_8^{-1/2}\left(dx_5^2+\cdots+dx_8^2\right)+\left(H_4H_8\right)^{1/2}ds_9^2\,,
\end{equation}
where the harmonic functions satisfy
\begin{equation}
\partial_{x_9}^2H_4+H_8\sum_{i=5}^8\partial_{x_i}^4H_4\,=\,0\,, \qquad \partial_{x_9}^2H_8\,=\,0\,.
\end{equation}
The solutions are given by
\begin{equation}
H_4\,=\,1+Q_4\left(r^2+\frac{4}{9}Q_8|x_9|^3\right)^{-5/3}\,, \qquad H_8\,=\,Q_8|x_9|\,,
\end{equation}
where $r^2\,=\,\sum_{i=5}^8(x_i)^2$ is the radial coordinate in the $x_5$ to $x_8$ directions and $Q_4$ and $Q_8$ are constant.

We specialized to the D4-branes smeared over the $x_5$ to $x_8$ directions. Thus the equations are 
\begin{equation}
\partial_{x_9}^2H_4\,=\,0\,, \qquad \partial_{x_9}^2H_8\,=\,0\,,
\end{equation}
and they are solved by
\begin{equation}
H_4\,=\,1+Q_4|x_9|\,, \qquad H_8\,=\,Q_8|x_9|\,.
\end{equation}
Near the core of the solution, $|x_9|\rightarrow{0}$, the metric is given by
\begin{equation}
ds^2\,=\,\left(Q_4Q_8\right)^{-1/2}|x_9|^{-1}\left(-dx_0^2+\cdots+dx_4^2\right)+Q_4^{1/2}Q_8^{-1/2}\left(dx_5^2+\cdots+dx_8^2\right)+\left(Q_4Q_8\right)^{1/2}|x_9|dx_9^2\,.
\end{equation}
We employ a change of coordinate,
\begin{equation}
x_9\,=\,r^{4/3}\,,
\end{equation}
and the metric reduces to
\begin{equation}
ds^2\,=\,\left(Q_4Q_8\right)^{-1/2}r^{-4/3}\left(-dx_0^2+\cdots+dx_4^2\right)+Q_4^{1/2}Q_8^{-1/2}\left(dx_5^2+\cdots+dx_8^2\right)+\left(Q_4Q_8\right)^{1/2}\frac{16}{9}r^2dr^2\,.
\end{equation}
This precisely matches the topological disk solution in the limit of $r\rightarrow{0}$ in \eqref{smeared48}.

%%%%%%%%%%%%%%%%%%%%%%%%%%%%%%%%%%%%%
\section{Equivalence with the spindle}
\renewcommand{\theequation}{C.\arabic{equation}}
\setcounter{equation}{0} 
%%%%%%%%%%%%%%%%%%%%%%%%%%%%%%%%%%%%%

Although the spindle and disk solutions are physically distinct, the solutions originate from different global completions of common local solutions. In this appendix, we show that the solution of topological disk we obtained in \eqref{metmet} matches the local solution in \cite{Faedo:2021nub}, by simple change of a coordinate, \eqref{simco}: by identifying the $r$ coordinate here with the scalar field, $X(y)$, in \cite{Faedo:2021nub}.

The spindle solution in \cite{Faedo:2021nub} is
\begin{align}
ds^2\,=&\,\left(y^2h_1h_2\right)^{1/4}\left(ds_{AdS_4}^2+\frac{y^2}{F}dy^2+\frac{F}{h_1h_2}dz^2\right)\,, \notag \\
A_i\,=&\,\left(\alpha_i-\frac{y^3}{h_i}\right)dz\,, \qquad X_i\,=\,\left(y^2h_1h_2\right)^{3/8}h_i^{-1}\,, \notag \\
F(y)\,=&\,m^2h_1h_2-y^4\,, \qquad h_i(y)\,=\,\frac{2\tilde{g}}{3m}y^3+q_i\,.
\end{align}
where $q_i$ and $\alpha_i$, $i\,=\,1,2$, are constants. We consider a special case of
\begin{equation}
\tilde{h}(y)\,\equiv\,h_1(y)\,=\,h_2(y)\,, \qquad X(y)\,\equiv\,X_1(y)\,=\,X_2(y)\,, \qquad q\,\equiv\,q_1\,=\,q_2\,, \qquad \alpha\,\equiv\,\alpha_1\,=\,\alpha_2\,.
\end{equation}
It is equivalent to the reduction of matter coupled $F(4)$ to pure $F(4)$ gauged supergravity.

Now we perform a change of coordinate by identifying the scalar field, $X(y)$, with the coordinate, $r$, in \eqref{ansatzz},
\begin{equation} \label{simco}
X(y)\,=\,r\,.
\end{equation}
Note that $y$ is the spindle coordinate of \cite{Faedo:2021nub} and $r$ is the disk coordinate of ours. We further make identifications of parameters,
\begin{equation}
\tilde{g}\,=\,\frac{1}{2}g\,,\qquad q\,=\,\frac{1}{6}b\,, \qquad \alpha\,=\,2m\,.
\end{equation}
Then the solution reduces to our solution obtained in \eqref{metmet} with the parameters,
\begin{equation}
g\,=\,1\,, \qquad \mathcal{C}\,=\,-1\,.
\end{equation}
This shows that our disk solution matches the local solution of spindle in \cite{Faedo:2021nub}.

\vspace{2cm}

\bibliographystyle{JHEP}
\bibliography{20210809}

%%%%%%%%%%%%%%%%%%%%%%%%%%%%%%%%%%%%%%%%%%%%%%%%%%%%%%%%%%%%
\end{document}